\documentclass[12pt]{article}
\usepackage{epsfig}
\usepackage{psfrag}
\usepackage{latexsym}
\usepackage{indentfirst}
\usepackage{fancyhdr}
\usepackage{amssymb}
\usepackage{amsmath}
\usepackage{amsfonts}
\usepackage{cite}
\usepackage{bbold}
\usepackage{color}
\usepackage[footnotesize]{caption2}
\usepackage{graphicx}
\usepackage[center,footnotesize,hang]{subfigure}
\usepackage{url}

\makeatletter

\newcommand{\Rmnum}[1]{\expandafter\@slowromancap\romannumeral #1@}
\makeatother

\newcommand{\pa}{\partial}

\def\be{\begin{equation}}
\def\ee{\end{equation}}
\def\bc{\begin{center}}
\def\ec{\end{center}}
\def\bea{\begin{eqnarray}}
\def\eea{\end{eqnarray}}

\catcode`@=11
\def\marginnote#1{}
\newcount\hour
\newcount\minute
\newtoks\amorpm
\hour=\time\divide\hour by60
\minute=\time{\multiply\hour by60 \global\advance\minute by-\hour}
\edef\standardtime{{\ifnum\hour<12 \global\amorpm={am}%
        \else\global\amorpm={pm}\advance\hour by-12 \fi
        \ifnum\hour=0 \hour=12 \fi
        \number\hour:\ifnum\minute<10 0\fi\number\minute\the\amorpm}}
\edef\militarytime{\number\hour:\ifnum\minute<10 0\fi\number\minute}
\def\draftlabel#1{{\@bsphack\if@filesw {\let\thepage\relax
   \xdef\@gtempa{\write\@auxout{\string
      \newlabel{#1}{{\@currentlabel}{\thepage}}}}}\@gtempa
   \if@nobreak \ifvmode\nobreak\fi\fi\fi\@esphack}
        \gdef\@eqnlabel{#1}}
\def\@eqnlabel{}
\def\@vacuum{}
\def\draftmarginnote#1{\marginpar{\raggedright\scriptsize\tt#1}}
\def\draft{\oddsidemargin 0.0truein
        \def\@oddfoot{\sl preliminary draft \hfil
        \rm\thepage\hfil\sl\today\quad\militarytime}
        \let\@evenfoot\@oddfoot \overfullrule 3pt
        \let\label=\draftlabel
        \let\marginnote=\draftmarginnote
   \def\@eqnnum{(\theequation)\rlap{\kern\marginparsep\tt\@eqnlabel}%
\global\let\@eqnlabel\@vacuum}  }
\catcode`@=12
\begin{document}
\title{\hskip4.3in{\normalsize \bf USTC-ICTS-11-13}\\{\bf Superluminal Neutrinos from Special Relativity with de Sitter Space-time Symmetry}}

\author{Mu-Lin Yan\footnote{E-mail address: mlyan@ustc.edu.cn},   Neng-Chao Xiao\footnote{E-mail address: ncxiao@ustc.edu}, Wei Huang\footnote{E-mail address: weihuang@mail.ustc.edu.cn}, Sen Hu\footnote{E-mail address: shu@ustc.edu.cn},
\\Wu Wen-Tsun Key Lab of Mathematics of Chinese Academy of Sciences\\
Department of Modern Physics and School of Mathematical Sciences\\
University of Science and Technology of China, Hefei, Anhui 230026, China
}
\maketitle

\abstract{We explore the recent OPERA experiment of superluminal neutrinos in the framework of Special Relativity with de Sitter space-time symmetry (dS-SR). According to Einstein a photon is treated as a massless particle in the framework of Special Relativity. In Special Relativity (SR) we have the universal parameter $c$, the photon velocity $c_{photon}$ and the phase velocity of a light wave in vacuum $c_{wave}=\lambda\nu$. Due to the null experiments of Michelson-Morley we have $c=c_{wave}$. The parameter $c_{photon}$ is determined by the Noether charges corresponding to the space-time symmetries of SR. In Einstein's Special Relativity (E-SR) we have $c=c_{photon}$. In dS-SR, i.e. the Special Relativity with $SO(4,1)$ de Sitter space-time symmetry, we have $c_{photon}>c$. In this paper, the OPERA datum are examined in the framework of dS-SR. We show that OPREA anomaly is in agreement with the prediction of dS-SR with $R\simeq 1.95\times 10^{12}l.y.$  Based on the $p$-$E$ relation of dS-SR, we also prove that the Cohen and Glashow's argument of possible superluminal neutrino's Cherenkov-like radiation is forbidden. We conclude that OPERA and ICARUS results are consistent and they are explained in the dS-SR framework.

}

\section{Introduction}\label{sec:intr}

The OPERA collaboration recently reported the evidence of superluminal behavior for muon neutrinos
$\nu_{\mu}$ with energies of a few tens of GeVs\cite{OPERA}. The arrival time of the  $\nu_{\mu}$
neutrino with average energy of 17 GeV is earlier by $\delta t =(60.7\pm6.9_{stat}\pm 7.4_{sys}) $ ns.
This translates into a superluminal propagation velocity for neutrinos by a relative amount
\begin{eqnarray}\label{fasterthanlight}
 \delta c_{\nu}& = & \frac{v_{\nu}-c}{c} =(2.48 \pm 0.28_{stat} \pm 0.30_{sys})\times 10^{-5}
\end{eqnarray}
with significance level of $6 \sigma$. This result is consistent with the earlier MINOS experiment
\cite{Minos} and FERMILAB79 experiment \cite{Fermilab79}.

This would be the most significant discovery in fundamental physics over the last several decades because
OPERA datum definitely indicates $v_\nu >c$. It challenges the Einstein's Special Relativity (E-SR) directly.
It is well known that E-SR has been the cornerstone of modern physics which is well-established by innumerable
experiments and observations. An outstanding feature of E-SR is a universal upper limit of speed,
namely the speed of light $c$ in vacuum. It is surprising that this limit of speed is broken by the OPERA experiment.
Furthermore, Cohen and Glashow \cite{CG} argued that, in the frame work of E-SR,
such superluminal neutrinos should lose energy by producing $e^+e^-$ pairs, through $Z^0$ mediated processes
analogous to Cherenkov radiation. Soon after their work, the ICARUS Collaboration reported that there was no such  energy loss signals that were observed \cite{ICARUS}. The OPREA and ICARUS experiments indicate that
it is time to re-examine the underlying basis of Special Relativity.
In this paper we attempt to solve the puzzles arisen from the OPREA and ICARUS experiments
in the framework of Special Relativity with de Sitter space-time symmetry (dS-SR) \cite{look,Lu74,Ours}.

As a fundamental theory Special Relativity (SR) is a theory on global space-time symmetry.
Such symmetry is the foundation upon which the whole physics is built.
It is well known that the space-time metric in E-SR is $\eta_{\mu\nu}=diag\{+,-,-,-\}$.
The most general transformation to preserve the metric $\eta_{\mu\nu}$ is the global Poincar\'e group
(or inhomogeneous Lorentz group $ISO(1,3)$). It is well known also that the
Poincar\'e group is the limit of the de Sitter group with radius of the pseudo-sphere
$R\rightarrow \infty$. A natural question arisen is whether there exists another type of de Sitter
transformation with $R$ finite which also leads to a
special relativity theory. In 1970's, K.H. Look (Qi-Keng Lu) and his
collaborators Z.L. Zou and H.Y. Guo pursued this problem and they got a highly nontrivial
positive answer. They succeeded in formulating the mathematic structure of the Special Relativity
with global de Sitter space-time symmetry \cite{look,Lu74}.
To the best of our knowledge, ref. \cite{Lu74} is the first publication to explore SR theory
in the framework of de Sitter space-time symmetry, i.e., dS-SR. In 2005, Yan, Xiao, Huang, Li \cite{Ours}
performed Lagrangian-Hamiltonian formulism for dS-SR with two universal constants $c$ and $R$, and
suggested the quantum mechanics of dS-SR. Ref. \cite{Ours} is the base of our investigation in the present paper.
During the past decade the theory were extensively discussed \cite{Guo1,Guo2,yan1,guo3,yan2,yan3}.

A meaningful and deep physical question is that what is the space-time symmetry for the real world?
E-SR is the limit of dS-SR with $R\rightarrow \infty$. People suspect that E-SR may be an approximation
of dS-SR with large enough $R$. To get the answer, one should pursue the physical effects beyond E-SR.
We will see that the OPERA anomaly is an experiment to determine the space-time symmetry of the real physical world.

In this paper, we accept Einstein's hypotheses that a photon can be treated as a massless particle,
and its velocity $c_{photon}$ is the physical propagating speed of a photon in vacuum. We do not assume
$c_{photon}=c$ beforehand.
The phase velocity of a light wave in vacuum is $c_{wave}=\lambda\nu$. The relationships
between $c,  c_{photon}$ and $c_{wave}$ in both E-SR and dS-SR are carefully studied in the paper.
In SR the universal parameter $c$ is required to be independent of the reference systems.
The famous null experiments of Michelson-Morley show that the phase velocity of a light wave
$c_{wave}=\lambda\nu$ is independent of the reference systems with very high accuracy.
Thus, Einstein's outstanding assumption of $c=c_{wave}$ is sound and it is the foundation of both E-SR and dS-SR.
What is new in this paper is that the $c_{photon}$ is derived from the Noether chargers generated from
the SR's space-time symmetries. We will reveal in the paper that $c_{photon}=c=c_{wave}$ for E-SR, but for dS-SR, we have $c_{photon}> c=c_{wave}$ . This is an interesting result follows from
the space-time symmetry, and there are no {\it ad hoc} considerations that are involved.
Since $m_\nu$ is rather small, it is easy to achieve conclusion of $c_{photon} > v_\nu >c$ when $E_\nu$
is large enough. Furthermore, the calculations based on dS-SR dispersion relation show that
the Cherenkov-like process of $\nu_\mu\rightarrow \nu_\mu+e^++e^-$ is forbidden. Consequently
the OPERA anomaly on superluminal neutrinos is well interpreted by means of dS-SR in the present paper.

The paper is organized as follows. In section \ref{sec:2} we briefly review the Special Relativity with
de Sitter space-time symmetry in terms of the Lagrangian-Hamiltonian formulism.
In section \ref{sec:3}, we analyze meanings of the universal parameter $c$
and the photon velocity $c_{photon}$ both in E-SR and in dS-SR.
In section \ref{sec:4}, we analyze the OPERA data in dS-SR. We find that OPERA anomaly
is in agreement with the prediction of $c$-$c_{photon}$ degeneracy breaking from dS-SR.
In section \ref{sec:5}, we discuss the Cohen-Glashow arguments, and show that there is no conflict
between ICARUS data and OPERA anomaly in the scenario of dS-SR.
Finally, we briefly discuss the conclusions reached in this paper.

\section{The Special Relativity with de Sitter Space-time Symmetry}\label{sec:2}

\begin{figure}[ht]
\begin{center}
\includegraphics[width=0.7\textwidth]{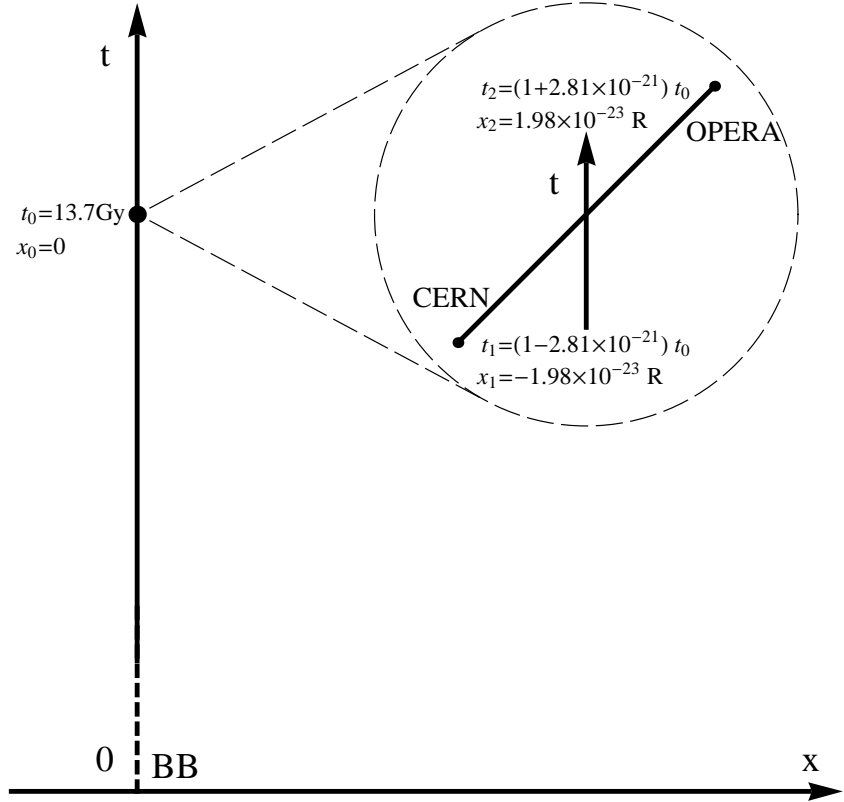}
\caption{\label{bb} Sketch of Natural Cosmical Reference System (NCRS):
The origin of NCRS is $(t,\;\mathbf{x})=(0,\;0)$, when and where the Big Bang occurred.
The picture of the OPERA experiment is sketched. In the dashed-circle the enlarged detail is shown.
The distance between CERN and OPERA (located in GranSasso) is about $\sim 731\;km$.
In the NCRS, the neutrino departing time and position are $t_1=(1-2.81\times 10^{-21})t_0$
and $x_1=-1.98\times 10^{-23}R$ respectively, and the arrival time and position are
$t_2=(1+2.81\times 10^{-21})t_0$ and $x_2=1.98\times 10^{-23}R$,
where $t_0=13.7Gy,\;R\simeq 1.95\times 10^{12}ly$ (see eq. (\ref{R})).
This figure shows that $t_1\simeq t_2\simeq t_0$ and $x_1/R\simeq x_2/R \simeq x_0/R\simeq 0$
are good approximations with high accuracy, which lead to the desired expressions of conserved Noether charges
for OPERA neutrinos in dS-SR from eq. (\ref{physical momenta}): $K^i\simeq -t_0p^i_{dS}$ and $L^i_{dS}\simeq 0$.}
\end{center}
\end{figure}

According to ref. \cite{Ours}, the Lagrangian for a free particle in dS-SR reads
\begin{equation}\label{271}
 L_{{dS}}(t,x^i,\dot{x}^i)=-m_0c\frac{ds}{dt}
 =-m_0c \frac{\sqrt{B_{\mu\nu}(x)dx^\mu dx^\nu}}{dt}
 =-m_0c \sqrt{B_{\mu\nu}(x)\dot{x}^\mu \dot{x}^\nu},
 \end{equation}
where $\dot{x}^\mu=\frac{d}{dt}x^\mu$, and the Beltrami metric $B_{\mu\nu}(x)$ serves as the inertial frame system in dS-SR, which is as follows
\begin{eqnarray}\label{star281}
B_{\mu\nu}(x)&=& \frac{\eta_{\mu\nu}}{\sigma (x)}
+\frac{\eta_{\mu\lambda}\eta_{\nu\rho} x^\lambda x^\rho}{R^2 \sigma(x)^2}, \\
\sigma(x) & \equiv & 1-\frac{1}{R^2}
\eta_{\mu\nu}x^\mu x^\nu, ~~
\eta_{\mu\nu} =diag(1,-1,-1,-1),
\end{eqnarray}
where the universal light-speed parameter $c$ and the radius $R$ of the pseudo-sphere
in de Sitter space are two universal constants in the theory. We would like to address two issues in the following:

\begin{enumerate}

\item From the principle of least action
\begin{equation}\label{272}
 \delta S\equiv \delta\int L_{{dS}}(t,x^i,\dot{x}^i)dt=0
 \end{equation}
we have \cite{Ours}
\begin{equation}\label{273}
 v^i=\dot{x}^i=constant.
 \end{equation}
 This is a highly non-trivial result since it indicates that just like $\eta_{\mu\nu}$ in E-SR the metric $B_{\mu\nu}(x)$ is indeed the inertial frame metric even though
the Lagrangian $L_{{dS}}(t,x^i,\dot{x}^i)$ deduced from $B_{\mu\nu}(x)$ is space-time dependent. The existence of inertial frame metric is a first-principle requirement of a SR theory, and hence $B_{\mu\nu}(x)$'s existence
means that E-SR is not the unique SR theory. dS-SR is its natural extension which is another candidate of SR. Since $B_{\mu\nu}(x)|_{(|R|\rightarrow \infty)}=\eta_{\mu\nu}$, we may think that (mechanics of dS-SR)$_{(|R|\rightarrow \infty)}=$ (mechanics of E-SR).
There is no any {\it prior} reason to assume that the space-time symmetry of the real world is as in E-SR or in dS-SR. We will show that the OPREA results favor dS-SR. It is significant to determine the magnitude of $R$ for the real world. In this paper we use the OPERA data to estimate $R$.

\item In dS-SR the space-time transformations are:
\begin{eqnarray}\label{transformation}
x^{\mu} \;-\hskip-0.10in\longrightarrow\hskip-0.3in^{dS}
 ~~~ \tilde{x}^{\mu} &=& \pm \sigma(a)^{1/2} \sigma(a,x)^{-1}
(x^{\nu}-a^{\nu})D_{\nu}^{\mu}, \\
    \nonumber D_{\nu}^{\mu} &=& L_{\nu}^{\mu}+R^{-2} \eta_{\nu
\rho}a^{\rho} a^{\lambda} (\sigma
(a) +\sigma^{1/2}(a))^{-1} L_{\lambda}^{\mu} ,\\
\nonumber L : &=& (L_{\nu}^{\mu})\in SO(1,3), \\
\nonumber \sigma(x)&=& 1-\frac{1}{R^2}{\eta_{\mu \nu}x^{\mu} x^{\nu}}, \\
\nonumber \sigma(a,x)&=& 1-\frac{1}{R^2}{\eta_{\mu \nu}a^{\mu} x^{\nu}}.
\end{eqnarray}

The metric tensor $B_{\mu\nu}(x)$ and the action of
dS-SR (\ref{271}) transform respectively as follows \cite{Lu74}\cite{Ours}
\begin{equation} \label{B01}
 B_{\mu\nu}(x)\;-\hskip-0.10in\longrightarrow\hskip-0.3in^{dS}
 ~~~ ~\widetilde{B}_{\mu\nu}(\widetilde{x})={\pa x^\lambda \over \pa
 \widetilde{x}^\mu}{\pa x^\rho \over \pa
 \widetilde{x}^\nu}B_{\lambda\rho}(x)=B_{\mu\nu}(\widetilde{x}),
\end{equation}
\begin{equation} \label{B02}
 S_{dS}\equiv\int dt L_{dS}(t,x^i,\dot{x}^i)=-m_0c\int dt {\sqrt{B_{\mu\nu}(x)dx^\mu dx^\nu} \over dt}
 \;-\hskip-0.05in\hskip-0.05in\longrightarrow\hskip-0.3in^{dS}
 ~~ ~~\widetilde{S}_{dS}=S_{dS}.
\end{equation}
There are ten parameters in the dS transformation of eq. (\ref{transformation}). The action is invariant
under those transformations. Hence there are 10 conserved Noether charges in dS-SR similar as in E-SR. They are as follows: \cite{Ours}
\begin{eqnarray}\nonumber
p_{dS}^i &=& m_0 \Gamma \dot{x}^i
\\
\nonumber  E_{dS} &=&  m_0 c^2 \Gamma
\\
\label{physical momenta} K_{dS}^{i} & =& m_0 c \Gamma (x^i -t\dot{x}^i)=m_0c\Gamma x^i-tp_{dS}^i
\\
\nonumber L_{dS}^{i} & =& -m_0 \Gamma \epsilon^{i}_{\;jk} x^j
\dot{x}^k=-\epsilon^{i}_{\;jk} x^jp^k_{dS}.
\end{eqnarray}

\noindent Here $E_{dS},{\mathbf p}_{dS},{\mathbf L}_{dS},{\mathbf K}_{dS}$ are
conserved physical energy, momentum, angular-momentum and boost
charges respectively, and $ \Gamma $ is:
 {\begin{eqnarray} \nonumber
 \Gamma^{-1}\hskip-0.1in =\sigma(x) \frac{ds}{c dt}=\frac{1}{R} \sqrt{(R^2-\eta_{ij}x^i
x^j)(1+\frac{\eta_{ij}\dot{x}^i \dot{x}^j}{c^2})+2t \eta_{ij}x^i
\dot{x}^j -\eta_{ij}\dot{x}^i \dot{x}^j t^2+\frac{(\eta_{ij}
x^i\dot{x}^j)^2}{c^2}}.\\ \label{new parameter}
\end{eqnarray}}
When $|R| \to \infty$, $\Gamma$ becomes the Lorentz factor $\gamma$.
 From these definitions, it is straightforward to check the identity of $\sigma^2(x)B_{\mu\nu}(x)p_{dS}^\mu p_{dS}^\nu=m_0^2c^2$. We then have the dispersion relation for dS-SR as follows \cite{Ours}
\begin{eqnarray}\label{dp}
E_{dS}^2 =m_0^2 c^4+{\mathbf p}_{dS}^2 c^2 + \frac{c^2}{R^2}
({\mathbf L}_{dS}^2-{\mathbf K}_{dS}^2).
\end{eqnarray}
(This relation were also suggested in \cite{Lu74} \cite{plus1}).
When $|R|\rightarrow \infty$, the above relation reduces to the well known dispersion relation in E-SR
\begin{eqnarray}\label{dp1}
E_{E}^2 =m_0^2 c^4+{\mathbf p}_{E}^2 c^2,
\end{eqnarray}
where $E_E=E_{dS}|_{(|R|\rightarrow\infty)},\;{\mathbf p}_E={\mathbf p}_{dS}|_{(|R|\rightarrow\infty)}$, and the subscript $E$ means E-SR's.
Comparing eq. (\ref{dp1}) with eq. (\ref{dp}), we see that the E-SR's dispersion relation is independent of the Noether charges of Lorentz boost
and rotations in space (angular momenta), and hence it is independent of the space-time coordinates origins of the reference frames (see eq. (\ref{physical momenta})).
However, in dS-SR's it is not. In other words, the dS-SR's dispersion relation (\ref{dp}) depends on a choice of  the space-time origin.
In the real world, the Big Bang (BB) cosmology model are widely accepted. In this model, the BB occurrence
provides a natural space-time coordinates origin. Then the current experiments in the Earth laboratory, e.g., OPERA measurements, are at the time $t_0\simeq 13.7Gy$ and $\mathbf{x}_0\equiv \mathbf{x}(t_0)\simeq 0$ (see figure \ref{bb}). In the follows, we will call this coordinates system the Natural Cosmical Reference System (or shortly NCRS). To NCRS, the time $t=0$ (i.e., the time-coordinate's origin) is the starting point of arrow of the cosmic time. For any positions on the isochronic hypersurface of NCRS, the corresponding 3D space is isotropous
and homogenous, i.e., the Copernicus principle of cosmology holds.
\end{enumerate}

\section{The Universal Parameter $\mathbf{c}$ and the
photon velocity$\mathbf{c_{photon}}$ in SR Mechanics}\label{sec:3}
It is well known that a Special Relativity theory (SR) is built on a 4-dimensional space-time manifold with a metric. Therefore, an Uniform Parameter $c$ (UP $c$) is needed to convert the time dimension to the length dimension. The UP $c$ has the same dimension as velocity and it is independent of a choice of the reference systems. Namely $c$ serves as an absolute velocity in special relativity theories. It is long known that the null experiments of Michelson and Morley \cite{MM} indicate that the phase velocity of a light wave $c_{wave}=\lambda\nu$ is independent of the reference systems with very high accuracy. Therefore, Einstein specifically assumed $c=c_{wave}$. This is the foundation of SR (for both E-SR and dS-SR). $c_{wave}=\lambda\nu$ can be directly measured by measuring
a beam of laser light's wavelength $\lambda$ and its frequency $\nu$ respectively.
Actually, in 1972 \cite{Speed}, the frequency and the wavelength of a beam of methane-stabilized laser at $3.39 \mu m$ were directly measured. With infrared frequency synthesis techniques, one obtained $\nu=88,376 181627(50)$ THz. With frequency-controlled interferometry, the authors of \cite{Speed} found $\lambda =3.392231376(12)\mu {\rm m}$. Multiplication yields the phase velocity of a light wave $c_{wave} = 299 792456.2(1.1)$ m/sec. It improved the measurement of
$c_{\{wave\}}$ by two digits to the previously accepted value. After that, at the 1983 {\it Conference Gener des Poids et Mesures}, the following SI (System International) definition of meter was adopted:
\hskip0.1in {\it The meter is the length of the path traveled by light in vacuum during a time interval of $1/299\;792\;458$ of a second.}\hskip0.1in This means that $c=c_{wave}$ has been defined to be {\it exactly} $299\;792\;458$m/s.

We would like to address here that the experiments of Michelson and Morley cannot determine the real world's space-time symmetry. The real world could be of Poincar\'e or de Sitter. Note that Einstein hypothesis of $c=c_{wave}$ holds for both E-SR and dS-SR.

The phase velocity of a light wave could be different from a photon's moving velocity. Now let us derive a photon's moving velocity $c_{photon}\equiv \dot{x}|_{m=0}$ (or its Hamilton-Jacobi velocity) from the E-SR mechanics and the dS-SR mechanics respectively.

\begin{enumerate}

\item The E-SR case:

In E-SR we have the metric is invariant with respect to the Poincare\'e group. Consequently there are 10 conserved Noether charges in E-SR. The charges are (e.g., see {\it pp581-586} and {\it Part 9} in ref. \cite{Noether}):

\begin{eqnarray}\label{503a}
\begin{array}{rcl}
 &&{\rm
Charges\;for\;space-translations\;(momenta):}~~~  p_E^i=m_0 \gamma \dot{x}^i, \\
 &&{\rm Charge\;for\;time-translation\;(energy): }~
 E_E= m_0 c^2 \gamma=\sqrt{c^2p_E^ip_E^j(-\eta_{ij})+m_0^2c^4} \\
&&{\rm{ Noether}\;charges\;for\;Lorentz\;boost:\;} ~~
 K^i_E=m_0 \gamma c (x^i- t \dot{x}^i)=m_0\gamma cx^i-ctp_E^i, \\
&&{\rm Charges\;for\;rotations\;in\;space\;(angular \;momenta):}~~~
L_E^i = \epsilon^{i}_{jk}x^{j}p_E^{k}.
\end{array}
\end{eqnarray}

According to Einstein, the light can be treated as a bunch of massless particles (i.e., photons whose $m_0=0$).
Hence, from eq. (\ref{503a}) and Einstein hypotheses the photon velocity $c_{photon}$ reads

\begin{eqnarray}\label{19a}
c_{photon}\equiv \dot{x}|_{m_0=0}=\left.{p_Ec^2\over E_E}\right|_{(m_0=0)}=\left.{p_Ec^2\over \sqrt{c^2p_E^2+m_0^2c^4}}\right|_{(m_0=0)}=c.
\end{eqnarray}

where superscript $i$ of $c_{photon},\;x,\;p_E$ were omitted for simpleness.
Eq. (\ref{19a}) indicates that the mechanical speed of light in vacuum $c_{photon}$ in E-SR equal to UP $c$,
i.e., $c_{photon}$ and $c$ are degenerate. However, the physical meanings of $c$ and $c_{photon}$
are different even though their magnitudes are equal in E-SR. We will see below that this is a key
to understand the recent OPERA and ICARUS experiments.

In the following, we will show that such degeneracy of $c$-$c_{photon}$ is broken in the dS-SR.

\item The dS-SR case:

We will calculate photon's mechanical velocity $c_{photon}$ in dS-SR by means of the Noether chages (\ref{physical momenta}) and the dispersion relation (\ref{dp}) in dS-SR. From Einstein's hypotheses and eq. (\ref{physical momenta}), we have
\begin{eqnarray}\label{28}
c_{photon}=\dot{x}|_{m_0=0}=\left.{c^2p_{dS}\over E_{dS}}\right|_{m_0=0}.
\end{eqnarray}
In NCRF, from (\ref{dp}) and (\ref{physical momenta}), $E_{dS}$ reads
\begin{eqnarray}\label{29}
E_{dS} =\sqrt{m_0^2c^4+{\mathbf p}_{dS}^2 c^2 + \frac{c^2}{R^2}
({\mathbf L}_{dS}^2-{\mathbf K}_{dS}^2)}=\sqrt{m_0^2c^4+p_{dS}^2 c^2(1- \frac{c^2t_0^2}{R^2})},
\end{eqnarray}
where NCRF condition: $\{x^i=0,\;t=t_0=13.7Gy\}$ has been used (see figure \ref{bb} and its caption). Substituting eq. (\ref{29}) into (\ref{28}), we obtain
\begin{eqnarray}\label{30}
c_{photon}={c\over \sqrt{1-{c^2t_0^2\over R^2}}}\neq c.
\end{eqnarray}
 From eqs. (\ref{28})(\ref{29}), we see that $c_{photon}$ is the upper limit of speed in dS-SR within NCRF.
For de Sitter $SO(4,1)$ case, $R^2>0$, then $c_{photon}>c$. For anti de Sitter $SO(3,2)$ case, $R^2<0$, then $c_{photon}<c$. Therefore, when $|R|\neq \infty$, the $c$-$c_{photon}$ degeneracy is broken in dS-SR.
\end{enumerate}

Finally, we briefly discuss the above results. We have shown that $c_{photon}=c=c_{wave}$ for E-SR, however, $c_{photon}\neq c=c_{wave}$ for dS-SR. Obviously this statement does not effect the existed SI standard for measurements in physics.
Thus, when we say there exist superluminal neutrinos in the framework of dS-SR, it means that the mechanical
velocity of neutrinos measured by the SI standard $v_\nu$ is greater than the $c\equiv 299\;792\;458$m/s. Physically,
$c=c_{wave}=\lambda\nu$ represents the phase velocity of a light wave in the vacuum, and $c_{photon}$ is the mechanical velocity of photons in vacuum. To directly determine $c_{photon}$ should be another OPERA-like experiment, in which one would use $\gamma$ with few GeV energies as moving particle (instead of the present OPERA's $\nu_\mu$). However it may not be an easy experiment. $R$ is another universal parameter in dS-SR. We will show in the following Section that OPERA anomaly provides a chance to determine the magnitude of $R$. When $R$ is known, of course, $c_{photon}$ shall be obtained via eq. (\ref{30}).

\section{Neutrinos Velocity from dS-SR}\label{sec:4}

The OPERA Collaboration reported that neutrino velocity is greater than
 $c$, the vacuum light speed in Einstein Special Relativity, by
 \begin{eqnarray}\label{fasterthanlight1}
 \delta c_{\nu}& = & \frac{v_{\nu}-c}{c} =(2.48 \pm 0.28_{stat} \pm 0.30_{sys})\times 10^{-5}
\end{eqnarray}
The velocity difference depends on energy slightly, by splitting the events into two groups with energies above or below 20 Gev, the velocity difference is given by
\begin{eqnarray}
  \delta c_{\nu} &=& (2.74\pm0.74\pm0.30)\times 10^{-5} {~~\rm{For \langle E \rangle=42.9 GeV}}\\
  \delta c_{\nu} &=& (2.16\pm0.76\pm0.36)\times 10^{-5} {~~\rm{For \langle E \rangle=13.9 GeV}}
\end{eqnarray}
The OPERA experiment is consistent with earlier experiments, as summarized in
Table \ref{table1} \cite{limiaowangyi},\cite{maboqiang}.
\begin{table}[htb]
\begin{center}
\caption{Summary of the neutrino velocity measurements}\label{table1}
\begin{tabular}{|c|c|c|c|c|}
 \hline \hline
 Experiment & Velocity  & Energy&  Flavors \\  \hline
  OPERA\cite{OPERA} & $\delta c_{\nu}=(2.48\pm0.28\pm 0.30)\times 10^{-5}$  & 17GeV & $\nu_{e},\bar{\nu}_{e},\nu_{\mu},\bar{\nu}_{\mu}$
  \\ \hline
  MINOS\cite{Minos} & $\delta c_{\nu}=(5.1\pm2.9)\times 10^{-5}$  & 3GeV & $\nu_{e},\bar{\nu}_{e},\nu_{\mu},\bar{\nu}_{\mu}$ \\   \hline
   FERMILAB79\cite{Fermilab79} &  $| \delta c_{\nu} | <4\times 10^{-5}$  & 3 $\sim$ 200GeV & $\nu$,$\bar{\nu}$ \\
  \hline
\end{tabular}
\end{center}
\end{table}

In dS-SR we consider neutrinos to be free moving massive point particles. Its velocity can be derived from the Noether charge formulars (\ref{physical momenta}).
Taking the OPERA neutrino moving trajectory as $\{x^1\equiv x(t), \;x^2=0,\; x^3=0\}$, from eqs. (\ref{physical momenta}) (\ref{dp}), we have:
\begin{eqnarray} \label{Energy1}
v_{dS}&\equiv &\dot{x}(t)=\frac{c^2 p_{dS}}{E_{dS}}, \\
\label{energy2} E_{dS} &=&\frac{m_0 c^2}{\sqrt{1-(\frac{v_{dS}}{c})^2+(\frac{x_0-v_{dS}t_0}{R})^2}},
\end{eqnarray}
where $t_0$ and $x_0$ are initial time and space location the OPERA moving neutrino's in NCRS, i.e., $t_0\simeq 13.7Gy,\;\;x_0=x(t_0)\simeq 0$.
 When $R \rightarrow  \infty$, it reduces to E-SR's
famous equation of $E=mc^2$. The last term of the denominator of the right side of eq. (\ref{energy2}) reflects the difference between dS-SR's dispersion relation and E-SR's, which comes from the term of ${c^2\over R^2}(\mathbf{L}_{dS}^2-\mathbf{K}_{dS}^2)$ of eq. (\ref{dp}) (note $\mathbf{L}_{dS}=0$ since $x^i=0$ here).  From eq. (\ref{energy2}) we have
\begin{eqnarray}\label{delta v}
  1-\frac{v_{dS}^2}{c^2} &=& \frac{m_0^2 c^4}{E_{dS}^2} -\frac{(x_0-v_{dS}t_0)^2}{R^2},
\end{eqnarray}
and then obtain the neutrino velocity
\begin{eqnarray}\label{19}
 v_{dS}=c \sqrt{1-\frac{m_0^2 c^4}{E_{dS}^2}\over 1-\frac{c^2 t_0^2}{R^2}}.
\end{eqnarray}
\begin{figure}[ht]
\begin{center}
\includegraphics[width=0.8\textwidth]{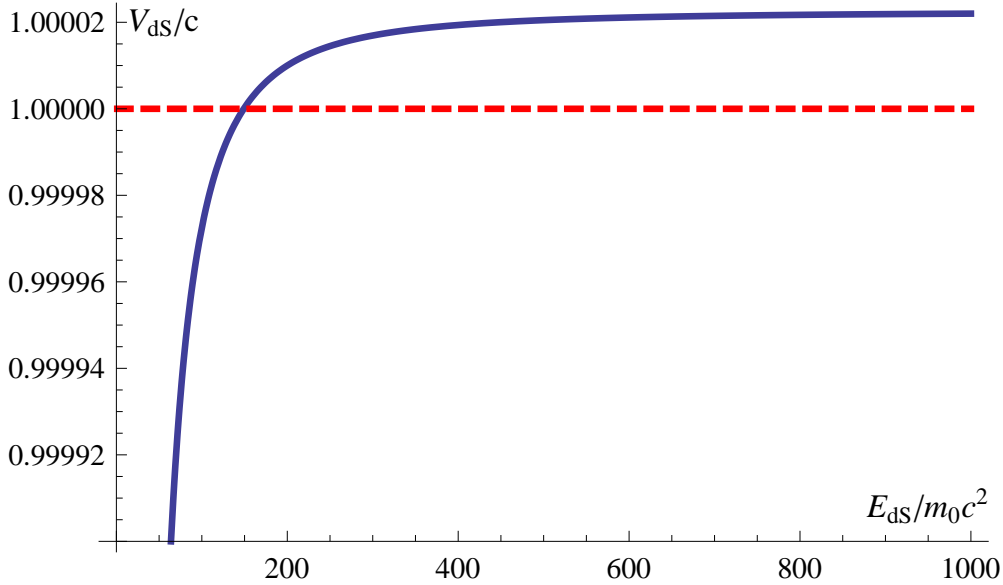}
\caption{Relation of velocity-energy of particles in dS-SR, eq. (\ref{19}).}\label{Energy}
\end{center}
\end{figure}
The function of $v_{dS}(E_{dS})$ is shown in figure \ref{Energy}.
From the figure we see that as the neutrino energy increases, the velocity increases. And such velocity-energy dependent is very weak when $E_{dS}>200m_0 c^2$. This is consistent with the OPERA's data \cite{A.Camelia} and FERMILAB79's data \cite{Fermilab79,maboqiang}.

It is easy to see that once the energy in eq. (\ref{19}) (or figure \ref{Energy}) is
greater than a critical value
\begin{eqnarray}
  E_{critical} &=&  \frac{R}{c t_0}m_0 c^2
\end{eqnarray}
 the neutrino velocity $v_{dS}$ in dS-SR will be greater than UP $c$.
That is the reason why OPERA measured superluminal signals. But when $E_{dS}<
 E_{critical}$, the velocity is smaller or equal to UP $c$, as SN
1987A neutrino data shows \cite{supernovabound}.

The parameter $R$ in dS-SR can be determined by the OPERA data of Eq.(\ref{fasterthanlight1}). Noting
$v_\nu=v_{dS}$, from Eqs. (\ref{fasterthanlight1}) and (\ref{19}), we have
\begin{equation}\label{19-1}
\delta c_\nu={v_{dS}-c\over c}=\sqrt{1-\frac{m_0^2 c^4}{E_{dS}^2}\over 1-\frac{c^2 t_0^2}{R^2}}-1
\simeq {c^2t_0^2\over 2R^2}-{m_0^2c^4\over 2E_{dS}^2}.
\end{equation}
To neutrinos from OPERA experiment, $E_{dS}\simeq 17GeV$, $m_0c^2\simeq 2eV$, the second term of
above equation (\ref{19-1})
 $m_0^2c^4/ 2E_{dS}\simeq 7\times 10^{-21}<< (\delta c_\nu)_{(OPERA)} \simeq 2.48\times 10^{-5}$. Hence this term can be ignored, and we have
\begin{equation}\label{19-2}
R\simeq {ct_0\over \sqrt{2\delta c_\nu}}
\simeq (1.95\pm 0.11\pm 0.12)
\times 10^{12} l.y,
\end{equation}
where OPERA datum (\ref{fasterthanlight1}) and $t_0=13.7Gy$ have been used.
This result is consistent with our previous result $0.45\times
10^{12}l.y.$\cite{yan1}, in which we tried to use Dirac large number hypotheses and dS-SR to
solve the inconsistence between the observational results of the QSO absorption lines and
of the Oklo nature reactor on the variation of the fine-structure constant.

\section{Exclusion of the Puzzle of Cherenkov-like Radiations of OPERA Neutrinos and ICARUS Data}\label{sec:5}

First, we would like to emphasize here that comparing with E-SR's, an outstanding feature of dS-SR's
Lagrangian-Hamiltonian formalism \cite{Ours} is that the canonical momenta-energy $(\vec{\pi},\;H)$ is not equal to conserved physical momenta-energy $(\mathbf{p},\;E)$. Therefore the Hamiltonian-Jacobi velocity (or group velocity)
\begin{equation}\label{HJ}
\mathbf{v}\equiv \mathbf{\dot{x}}={\partial H\over \partial {\vec{\pi}}}\neq {\partial E\over \partial {\mathbf{p}}},
\end{equation}
where the canonical momentum $\pi_{i} =\frac{\pa L_{dS}}{\pa
\dot{x}^i} = -m_0 \sigma(x) \Gamma B_{i \mu}\dot{x}^{\mu}\neq p^i_{dS}$, and the canonical energy (or Hamiltonian) $H
=\sum_{i=1}^3 \frac{\pa L_{dS}}{\pa \dot{x}^i} \dot{x}^i
-L_{dS}=m_0 c \sigma(x) \Gamma B_{0 \mu}\dot{x}^{\mu}\neq E_{dS}$ (see eq. (\ref{physical momenta})).
To the real world, $v=v_{dS},\;E=E_{dS},\;p=p_{dS}$ (hereafter the subscript ``dS'' will be omitted).
It can be proved straightforwardly in the Lagrangian-Hamiltonian formalism of dS-SR that the Hamiltonian-Jacobi velocity is as follows \cite{Ours}
\begin{equation}\label{HJ1}
\mathbf{v}\equiv \mathbf{\dot{x}}={\partial H\over \partial {\vec{\pi}}}={c^2\mathbf{p}\over E},
\end{equation}
which is the same as the expression of $\dot{\mathbf{x}}$ eq. (\ref{Energy1}) derived from the dS-SR Noether charge formula eq. (\ref{physical momenta})\cite{footnote}.

The physically measured quantities are $(\mathbf{v}\equiv \mathbf{\dot{x}},\;E,\;\mathbf{p})$ of particles.
Usually the phrase of superluminal particle means $v\equiv |\mathbf{v}|>c$. The relationship between $v$ and $(E,\;p)$ has been shown in eq. (\ref{Energy1}) (or eq. (\ref{HJ1})), i.e., $v={c^2p\over E}$. To the OPREA neutrino $\nu_\mu$, from eqs. (\ref{30}), (\ref{19}) and (\ref{19-1}), we have
\begin{eqnarray}\label{3v}
  c<v_\nu<c_{photon}.
\end{eqnarray}
Hence OPREA neutrinos are superluminal even though $v_\nu$ is less than the upper limit of speed of dS-SR $c_{photon}$.

In the framework of E-SR, Cohen and Glashow argued\cite{CG} that superluminal neutrinos should lose
energy through the $Z^0$ mediated process analogous to Cherenkov
radiation: $\nu_{\mu}\rightarrow \nu_{\mu}+e^{+}+e^{-}$. Cohen-Glashow model is based on a dispersion relation
\begin{equation}\label{CG1}
E^2=c^2p^2(1+\delta),~~~~with~~\delta>0,
\end{equation}
which breaks E-SR space-time symmetry. It is essential that one can determine whether the Cherenkov-like process occurs or not via examining the energy threshold of that process based on the dispersion relation violating the Lorentz invariance. In ref. \cite{mli}, the threshold of $\nu_{\mu}(p)\rightarrow \nu_{\mu}(p')+e^{+}(k')+e^{-}(k)$ has been derived as follows
\begin{equation}\label{CG2}
(E^2-p^2c^2)_{\rm thr.}=(2m_e+m_\nu)^2c^4.
\end{equation}
Substituting eq. (\ref{CG1}) into eq. (\ref{CG2}), we have \cite{mli}
\begin{equation}\label{CG3}
E_{\rm thr.}={(2m_e+m_\nu)c^2\over \sqrt{1-{1\over 1+\delta}}}\simeq {2m_ec^2\over \sqrt{\delta}},
\end{equation}
that is the same as the threshold in \cite{CG}. The existence of $E_{\rm thr.}>0$ indicates that the Cherenkov-like process $\nu_{\mu}(p)\rightarrow \nu_{\mu}(p')+e^{+}(k')+e^{-}(k)$ does occur when $E>E_{\rm thr.}$. The main conclusion of \cite{CG} were checked by means of the threshold equation (\ref{CG2}) and the model's dispersion relation (\ref{CG1}).

Now we view the dS-SR's $E$-$p$ relation of eq. (\ref{29}) as dispersion relation violating Lorentz symmetry, and copy it as follows
\begin{equation}\label{CG4}
p^2c^2={E^2-m_\nu^2c^4\over 1-{c^2t_0^2\over R^2}}.
\end{equation}
Substituting eq. (\ref{CG4}) into eq. (\ref{CG2}), we get
\begin{equation}\label{CG5}
E_{\rm thr.}^2=-{R^2\over c^2t_0^2}[(2m_e+m_\nu)^2c^4(1-{c^2t_0^2\over R^2})
-m_\nu^2c^4]\simeq -{R^2\over c^2t_0^2}4m_e^2c^4.
\end{equation}
Obviously, there is no real and positive solution of $E_{\rm thr.}$ from eq. (\ref{CG5}). Namely under the dispersion relation of eq. (\ref{CG4}), the threshold of process $\nu_{\mu}(p)\rightarrow \nu_{\mu}(p')+e^{+}(k')+e^{-}(k)$ is absent, and hence that Cherenkov-like process is forbidden kinematically. Similarly, since a photon's dispersion relation is the same as eq. (\ref{CG4}) except $m_\nu\Rightarrow m_\gamma=0$, eq. (\ref{CG5}) means also that the high energy photon's Cherenkov-like process $\gamma(p)\rightarrow \gamma(p')+e^{+}(k')+e^{-}(k)$ is also forbidden kinematically.

Consequently, we conclude that the Cohen-Glashow events caused energy loss for the superluminal neutrinos
are absent in the SO(4,1) dS-SR model. This is consistent with recent experiment result of the ICARUS collaboration,
another neutrino group in Gran Sasso\cite{ICARUS}. They reported no such energy spectrum shift signals were seen as predicted by Cohen-Glashow.

\section{Conclusions and Discussions}\label{sec:con}

Recently, the OPERA experiment of superluminal neutrinos has been widely discussed \cite{CG,mli,limiaowangyi,maboqiang,plus2,plus3,plus4,plus5,plus6,plus7}.
In the present paper we explore it in the framework of Special Relativity with de Sitter space-time symmetry (dS-SR).
Einstein's hypotheses that a photon can be treated as a massless particle in the Special Relativity are employed to define the kinematics of photons. The physical meanings of the universal parameter $c$, the photon velocity $c_{photon}$ and the phase velocity of a light wave $c_{wave}=\lambda\nu$ in E-SR and in dS-SR have been analyzed.
 By the null experiments of Michelson-Morley, $c$ can be conveniently taken to be $c=c_{wave}=\lambda\nu$, which can be determined by measuring the $\lambda$ and $\nu$ of lasers respectively. SI standard is available for both E-SR and dS-SR, and a massive particle with velocity $v > c$ is superluminal.

 The photon velocity $c_{photon}$ is determined by the Noether charges of SR. We found out that $c=c_{photon}$ in E-SR, yet $c\neq c_{photon}$ in dS-SR. This $c$-$c_{photon}$ degeneracy-breaking effect in dS-SR is an outstanding feature of dS-SR, which is implied by the space-time symmetry of dS-SR. For the $SO(4,1)$-de Sitter symmetry we have $c_{photon}>c$. Therefore in dS-SR it predicts that the velocity of a particle with zero or very small mass can be larger than the universal parameter $c$. Based on this analysis we examined the OPERA data and it is revealed that OPERA anomaly is in agreement with the prediction of $SO(4,1)$ dS-SR with $R\simeq 1.95\times 10^{12}l.y.$

 The Cohen and Glashow's argument on the possible superluminal neutrino's energy loss by producing  $e^+e^-$ pairs has also been discussed in detail. Starting from the dS-SR's $p$-$E$ relation (i.e., dispersion relation violating Lorentz invariant) for superluminal neutrinos we show that that such Cherenkov-like is forbidden kinematically. Due to such a forbidden mechanism the ICARUS collaboration did not see such sort of energy loss signals. It is consistent with the OPERA experiment in the framework of dS-SR. The conclusion reached in this paper is that the OPERA experiment and the ICARUS experiment are evidences to support that the global space-time symmetry is de-Sitter.

\begin{center} {\bf ACKNOWLEDGMENTS}
\end{center}
{The authors acknowledge Professor Miao Li and Tower Wang for discussions on the Cohen-Glashow process.
 This work is partially supported by
National Natural Science Foundation of China under Grant No.~10975128
and No.~11031005 and by the Wu Wen-Tsun Key Laboratory of Mathematics at USTC of Chinese Academy of Sciences.}
\medskip

{\bf Notes added:}

\medskip

After this work, we learned the following developments. On February 22, 2012 there was a report from ``Nature"'s web-site ``Flaws found in
faster-than-light neutrinos". It contains OPREA's official statement:

``The OPERA Collaboration, by continuing its campaign of verifications on the neutrino velocity measurement, has identified two issues that could significantly affect the reported result. The first one is linked to the oscillator used to produce the events time-stamps in between the GPS synchronizations. The second point is related to the connection of the optical fiber bringing the external GPS signal to the OPERA master clock. "

``These two issues can modify the neutrino time of flight in opposite directions. While continuing our investigations, in order to unambiguously quantify the effect on the observed result, the Collaboration is looking forward to performing a new measurement of the neutrino velocity as soon as a new bunched beam will be available in 2012. An extensive report on the above mentioned verifications and results will be shortly made available to the scientific committees and agencies."

It also reported: ``At Fermilab in Batavia, Illinois, members of the MINOS collaboration (Main Injector Neutrino Oscillation Search) continue to try to make their own independent measurement of the speed of neutrinos, with initial results expected later this year."

According to the analysis in this paper we would like to predict that new experiments would support
the main conclusion that Nature favors dS-SR and hence there are faster-than-light neutrinos.




\begin{thebibliography}{99}

\bibitem{OPERA}T. Adam et al.,
{\it Measurement of the neutrino velocity with the OPERA detector in the CNGS beam},
arXiv:1109.4897 [hep-ex].

\bibitem{Minos} P. Adamson et al., Phys.Rev.D {\bf 76}, 072005 (2007), arXiv:0706.0437 [hep-ex].

\bibitem{Fermilab79} G.R. Kalbfleisch, N. Baggett, E.C. Fowler and J. Alspector,
Phys. Rev. Lett. {\bf 43}, 1361 (1979);  J. Alspector et al., Phys. Rev. Lett. {\bf36}, 837 (1976).

\bibitem{CG} A.G. Cohen, S.L. Glashow, {\it New Constraints on Newtrino Velocity}, Phys. Rev. Lett., {\bf 107} 181803 (2011), arXiv:1109.6562 [hep-ph].


\bibitem{ICARUS} M. Antonello et al, {\it  Asearch for the analogue to Cherenkov radiation by high energy newtrinos at superluminal speeds in ICARUS }, arXiv: 1110.3763 [hep-ex].


\bibitem{look} K.H. Look (Q.K. Lu), {\it Why the Minkowski metric must be used ?}, (1970), unpublished.
\bibitem{Lu74} K.H. Look, C.L. Tsou (Z.L. Zou) and H.Y. Kuo (H.Y. Guo), {\it Acta Physica Sinica}, {\bf 23} (1974) 225 (in Chinese).

\bibitem{Ours} M.L. Yan, N.C. Xiao, W. Huang, S. Li, {\it Hamiltonian Formalism of the de-Sitter Invariant Special Relativity}, Commun.Theor.Phys.{\bf 48} (2007) 27, arXiv:hep-th/0512319.

\bibitem{Guo1} H.Y. Guo, C.G. Huang, Z. Xu, and B. Zhou, Phys. Lett. {\bf A331} (2004) 1;
Mod. Phys. Lett. {\bf A19} (2004) 1701; Chin. Phys. Lett. {\bf 22} (2005) 2477; arXiv:hep-th/0405137;
H.Y. Guo, C.G. Huang and B. Zhou, arXiv:hep-th/0404010.
\bibitem{Guo2} Y. Tian, H.Y. Guo, C.G. Huang, Z. Xu and B. Zhou, Phys. Rev. {\bf D71} (2005) 044030.

\bibitem{yan1}  S.X. Chen, N.C. Xiao, Mu-Lin Yan, {\it  Variation of the Fine-Structure Constant from the de Sitter Invariant Special Relativity}, Chinese Phys. {\bf C}, {\bf 32}, 612 (2008), arXiv:astro-ph/0703110.

\bibitem{guo3} H.Y. Guo, {\it Snyder's Model -- de Sitter Special Relativity Duality and de Sitter Gravity}, Class. Quant. Grav.{\bf 24}, 4009 (2007), arXiv:gr-qc/0703078.

\bibitem{yan2} M.L. Yan, {\it Hydrogen Atom and Time Variation of Fine-Structure Constant}, arXiv:1004.3023.

\bibitem{yan3}  M.L. Yan, Chinese Phys. {\bf C 35}, 228-232 (2011), arXiv: 1105.5693[physics.gen-ph].

\bibitem{supernovabound} M.J. Longo,  Phys. Rev. D {\bf 36}, 3276 (1987); L. Stodolsky, Phys. Lett. B
{\bf 201}, 353 (1988); G. F. Giudice, S. Sibiryakov and A. Strumia, arXiv:1109.5682.

\bibitem{mli} M. Li, D. Liu, J. Meng, T. Wang, L. Zhou, {\it ``Replaying neutrino bremsstrahlung with general dispersion relations''}, arXiv: 1111.3294 [hep-ph].


\bibitem{Gonzalez-Mestres} L. Gonzalez-Mestres, {\it Astrophysical consequences of the OPERA superlumial neutrino}, arXiv:1109.6630.

\bibitem{A.Camelia} G. Amelino-Camelia, G. Gubitosi, N. Loret, F. Mercati, G. Rosati, P. Lipari, {\it OPERA-reassessing data on the energy dependence of the speed of neutrinos}, arXiv:1109.5172.

\bibitem{plus1} T. Fukuyama, Ann. Phys. {\bf 157}, 321 (1984).

\bibitem{Noether}Edward A. Desloge. {\it Classical Mechanics},  John Wiley, New york, 1982.

\bibitem{MM} A. A. Michelson and E. W. Morley, Am. J. Sci. 34, S3S (1887); G. Joos, Ann. Phys. 7, 885 (1980); A. Brillet and J. L. Hall, Phys. Rev. Lett., {\bf 42}, 549 (1979).

\bibitem{Speed} K. M. Evenson, J. S. Welis, F. R. Petersen, B. I. Danielson, and G. W. Day,   Phys. Rev. Lett., {\bf 29}, 1346 (1972).

\bibitem{limiaowangyi} M. Li, Y. Wang, {\it Mass-dependent Lorentz Violation and Neutrino Velocity}, arXiv:1109.5924.

\bibitem{footnote}
The ``group velocity'' formula  $|{\mathbf v}|=|{\partial E\over \partial {\mathbf p}}|$ used usually is only available for the mechanics whose $\{H, |\pi_i| \}=\{E, |\mathbf{p}|\} $ where $\pi_i$ is canonical momentum and $\mathbf{p}$ is mechanical momentum. To the free particle's mechanics of E-SR, the usual ``group velocity'' formula holds. However, it does not hold for the mechanics of dS-SR. So, we cannot use the dispersion relation of eq. (\ref{29}) and the usual formula $v|_{group}={\partial E\over \partial p}$ to derive the particle's velocity in dS-SR mechanics. In this case, eq. (\ref{HJ1}) and (\ref{physical momenta}) are necessary. This is due to the requirement of the mechanics principle.

\bibitem{maboqiang} N. Qin, B.Q. Ma, arXiv:1110.4443; L.L. Zhou, B.Q. Ma, arXiv:1109.6097; L.L. Zhou, B.Q. Ma, arXiv:1109.6387; L.L. Zhou, B.Q. Ma, arXiv:1110.1850. N. Qin, B.Q. Ma, arXiv:1110.4443; L.L. Zhou, B.Q. Ma, arXiv:1109.6097; L.L. Zhou, B.Q. Ma, arXiv:1109.6387; L.L. Zhou, B.Q. Ma, arXiv:1110.1850.

\bibitem{plus2} J.W. Moffat, {\it Bimetric Relativity and the Opera Neutrino Experiment}, arXiv:1110.1330.

\bibitem{plus3} I. Oda and H. Taira, {\it A Resolution to Cherenkov-like Radiation of OPERA Neutrino}, arXiv:1110.6571.

\bibitem{plus4} C.A.G. Almeida, M.A. Anacleto, F.A. Brito, E. Passos. {\it Inflationary Cosmology and Superluminal Neutrinos}, arXiv:1112.0300 [hep-ph].

\bibitem{plus5} Bo-Qiang Ma, {\it The Phantom of the OPERA: Superluminal Neutrinos}, arXiv:1111.7050 [hep-ph].

\bibitem{plus6} S. Mohanty, S. Rao, {\it Constraint on super-luminal neutrinos from vacuum Cerenkov processes},  arXiv:1111.2725 [hep-ph].

\bibitem{plus7} F. Bezrukov, H.M.Lee, {\it Model dependence of the bremsstrahlung effects from the superluminal neutrino at OPERA}, arXiv:1112.1299 [hep-ph].

\end{thebibliography}
\end{document}